Hall effect of FeTe and Fe(Se$_{1-x}$Te$_x$) thin films

I. Tsukada[a,c,*], M. Hanawa[a,c], Seiki Komiya[a,c], A. Ichinose[a,c], T. Akiike[b,c], Y. Imai[b,c], A. Maeda[b,c]

a Central Research Institute fo Electric Power Industry, 2-6-1 Nagasaka, Yokosuka, Kanagawa 240-0196, Japan

b Department of Basic Science, The University of Tokyo, 3-8-1 Komaba, Meguro, Tokyo 153-8902, Japan

c Japan Science and Technology Agency, 5 Sanbancho, Chiyoda, Tokyo 102-0075, Japan

Abstract

The Hall effect is investigated in thin-film samples of iron-chalcogenide superconductors in detail. The Hall coefficient ($R_H$) of FeTe and Fe(Se$_{1-x}$Te$_x$) exhibits a similar positive value around 300 K, indicating that the high-temperature normal state is dominated by hole-channel transport. FeTe exhibits a sign reversal from positive to negative across the transition to the low-temperature antiferromagnetic state, indicating the occurrence of drastic reconstruction in the band structure. The mobility analysis using the carrier density theoretically calculated reveals that the mobility of holes is strongly suppressed to zero, and hence the electric transport looks to be dominated by electrons. The Se substitution to Te suppresses the antiferromagnetic long-range order and induces superconductivity instead.

The similar mobility analysis for Fe(Se$_{0.4}$Te$_{0.6}$) and Fe(Se$_{0.5}$Te$_{0.5}$) thin films shows that the mobility of electrons increases with decreasing temperature even in the paramagnetic state, and keeps sufficiently high values down to the superconducting transition temperature. From the comparison between FeTe and Fe(Se$_{1-x}$Te$_x$), it is suggested that the coexistence of `itinerant' carriers both in electron and hole channels is indispensable for the occurrence of superconductivity.

* Corresponding author

Dr. Ichiro Tsukada

Postal Address: Materials Science Research Laboratory

2-6-1 Nagasaka, Yokosuka, Kanagawa 240-0196 Japan

Phone: +81-46-856-2121

Fax: +81-46-856-5571

E-mail address: ichiro@criepi.denken.or.jp

1. Introduction

New iron-based superconductors have attracted much attention since LaFeAs($O_{1-x}F_x$) was reported to show superconductivity at $T$ = 26 K [1]. The highest $T_c$ ever reported reaches 56 K [2], and still great efforts have been made to raise their $T_c$'s. Iron-chalcogenide superconductors have the simplest structure among all the iron-based superconductors, and consist of only Fe and *Ch* (*Ch* = S, Se, and Te) [3,4]. There is a similarity in the crystallographic structure of Fe-*Ch* layer to that of Fe-*Pn* (*Pn* = As and P) layer implying a similar electronic state in both iron-pnictide and iron-chalcogenide superconductors. Indeed, the band calculation [5, 6] predicted that there are five bands mainly originated from five 3*d* levels of Fe crossing the Fermi level to form Fermi surface.

However, these two compounds have several intrinsic differences. In Fe*Pn*, explicit carrier doping is possible by chemical substitution by elements with a different valence, and a rigid-band picture properly explains the evolution of electronic states by carrier doping. In Fe*Ch*, however, superconductivity is induced by isovalent substitution of Se and/or S to Te in FeTe, which suggests that the rigid-band picture no longer works. Thus the evolution of electronic states from parent antiferromagnetic to doped superconducting states might be more complicated than Fe*Pn* superconductors, and hence the electric-transport measurements, especially Hall measurements, using high-quality single-crystalline samples are indispensable for understanding the evolution of electronic states in Fe*Ch*.

In this paper, we present the detailed results on a comparative study of Hall measurements in both parent antiferromagnet FeTe, and superconducting Fe(Se$_{1-x}$Te$_x$). We apply a semi-classical two-band Drude model and perform a phenomenological analysis of resistivity ($\rho$) and Hall coefficient ($R_H$) as functions not only of temperature but also of magnetic field. With the aid of carrier densities theoretically calculated for FeTe, we evaluate a mobility of electrons and holes, and discuss the key factor for the occurrence of superconductivity.

2. Experimental

Table I shows the sample specification. All the thin films were prepared by pulsed laser deposition method from carefully prepared polycrystalline target. Details are described elsewhere [7, 8]. We have selected several substrate materials suitable for thin-film growth of iron chalcogenide superconductors. In the present study, all the films are grown on MgO (100) or LaAlO$_3$ (100), which have been already confirmed to be appropriate for growing Fe(Se$_{1-x}$Te$_x$) [8]. We show the data of seven thin films: two FeTe, two Fe(Se$_{0.4}$Te$_{0.6}$), and three Fe(Se$_{0.5}$Te$_{0.5}$) thin films. One may notice that two Fe(Se$_{0.4}$Te$_{0.6}$) films are very thin (11nm), but their $c$-axis length are close to those reported for polycrystalline sample as those of the other relatively thick films are [9], and thus we expect no significant difference caused by the film thickness. We used a metal mask to make the film in a six-terminal shape as shown in Fig.1 in order not only to measure Hall resistance precisely but also to measure the thickness using a stylus profiler [10]. Longitudinal and transverse resistivities are measured using Physical Properties Measurement System (PPMS) under the magnetic field up to $\mu_0 H = 13$ T.

3. Results and Discussion

3.1 X-ray diffraction and transmission electron microscopy

Figure 2(a) shows x-ray diffractions of FeTe films. All the films have highly *c*-axis oriented structure. The calculated *c*-axis lengths are summarized in Table I. In both cases (MgO and LaAlO$_3$ substrates), the *c*-axis lengths of the films are comparable to that of Fe$_{1.07}$Te bulk crystal. This suggests that the films do not feel tensile stress in contrast to what is reported by Han *et al.* [11]. It should be also noted that the *c*-axis length is not so much different between the films on MgO and LaAlO$_3$, which is a similar result to what was observed in Fe(Se$_{0.5}$Te$_{0.5}$) thin films [8]. The substitution of Se shrinks the *c* axis as shown in Figs. 2(b) and 2(c). Fe(Se$_{0.4}$Te$_{0.6}$) and Fe(Se$_{0.5}$Te$_{0.5}$) films show the shorter *c*-axis length than FeTe. However, in our experiments, we did not see an explicit correlation of the chemical composition and the *c*-axis length between these two compounds.

Figures 3(a) and 3(b) show cross sectional images of FeTe and Fe(Se$_{0.5}$Te$_{0.5}$) thin films. In our previous report, we have revealed that the diffusion of oxygen to the grown film becomes significant on some substrate materials, such as YSZ and LaSrGaO$_4$. On MgO and LaAlO$_3$ substrates, however, the interface is quite sharp and no trace of oxygen diffusion is observed in both FeTe and Fe(Se$_{0.5}$Te$_{0.5}$) thin films. This property is quite beneficial for better in-plane orientation of the grown films. On both substrates, we always obtain MgO [100] ∥ Fe$Ch$ [100] and LaAlO$_3$ [100] ∥ Fe$Ch$ [100] as was also confirmed by x-ray diffraction, while on other substrates we frequently observe domains that have different in-plane orientations [8]. Therefore, we use MgO (100) and LaAlO$_3$ (100) for FeTe and Fe(Se$_{1-x}$Te$_x$) whenever we need to grow a `single-crystalline' Fe$Ch$ thin films.

3.2 Resistivity

The temperature dependence of resistivity ($\rho$) is summarized in Fig. 4. Let us first see the data of FeTe. The magnitude of resistivity is as low as that reported for bulk single crystals in both the films, while the details are different from the bulk crystals [12-14]. The most remarkable difference is the absence of discontinuous jump in $\rho$ indicating that a sharp first-order tetragonal-to-monoclinic structural transition does not occurs in these films. Instead a broad peak appears around 80 K, which may be due to the influence of epitaxy with the substrate. However, the resistivity behavior below 80 K is roughly the same as that of bulk crystals, and we may infer the antiferromagnetic long-range order evolves in the low-temperature phase.

Fe(Se$_{0.4}$Te$_{0.6}$) and Fe(Se$_{0.5}$Te$_{0.5}$) show superconductivity. The $T_c$ of S40T60-1 is 6.0 K, while that of S50T50-2 reaches 11.4K, which is the highest $T_c$ ever observed in our films. In both cases, we did not observe a significant influence of substrate materials to the resistivities. All the films show a similar $T$-dependence at relatively high temperature; $d\rho / dT$ is negative at room temperature, and then turns positive with decreasing temperature. The broad peak in $\rho$ is one of the common features of Fe$Ch$ superconductors, while is absent in FeTe.

In contrast to the `robust' $T$-dependence at high temperatures, resistivity behavior is rather scattered at low temperatures. As was discussed in Ref. [10], even the films with the same chemical composition can show either metallic ($d\rho / dT > 0$) or localizing ($d\rho / dT < 0$) behavior just above $T_c$. In our previous study, we reported that Fe(Se$_{0.5}$Te$_{0.5}$) films with relatively higher $T_c$ tend to show a metallic $T$-dependence. However, it has been revealed that it is not always the case. For example, as shown in Fig. 4, one of the Fe(Se$_{0.5}$Te$_{0.5}$) films (S50T50-3) shows very low resistivity and typical metallic behavior ($d\rho / dT > 0$) while its $T_c$ is as low as 3.4K. This result means that there is more than one factor determining $T_c$, and we need to analyze the normal-state transport properties in more quantitative manner, and understand the role of electron- and hole-channel conduction separately.

3.3 Hall effect

In order to reveal the complicated interplay between electrons and holes, we carried out Hall-effect measurements. Because the presence of anomalous Hall effect (AHE) was originally discussed in FeSe thin films [15], it is necessary to measure the magnetic-field dependence of transverse resistivity ($\rho_{xy}$), and to identify which field range is free from the influence of AHE. For that purpose, we first checked the field dependence of Hall resistivity ($\rho_H = [\rho_{xy}(H) - \rho_{xy}(-H)]/2$).

Figure 5(a) shows $\rho_H$ vs $\mu_0 H$ at 300 K for FeTe-1. In contrast to our previous studies, AHE is hardly observable in FeTe-1 (and also in FeTe-2). Nevertheless, we omit the data in the field range of -2 T < $H$ < 2 T, because the step-like behavior typical for AHE frequently can show up when the films orientation is not perfect [16], and use the rest of the data to calculate Hall coefficients ($R_H$). In case of Fe(Se$_{0.5}$Te$_{0.5}$), the field range of AHE is suppressed to -1 T < $H$ < 1 T as shown in Fig 5(b). The origin of the AHE has not been clarified yet. Unfortunately we could not measure spontaneous magnetization of the films simply because the sample volume is insufficient. Thus, we cannot compare the magnitude of anomalous term of Hall resistivity and magnetization. However, as was mentioned before, AHE does not always show up in FeTe and Fe(Se,Te) thin films, and also has been never reported in Fe(Se,Te) bulk crystals, which strongly indicates that AHE is not intrinsic to Fe*Ch* compounds.

The temperature dependence of $R_H$ is summarized in Fig. 5(c). In case of FeTe, $R_H$ shows a slight increase with decreasing temperature. Below 100 K, $R_H$ decreases rapidly, turns negative, and saturates at a negative constant value. This steep sign reversal is consistent with that reported in single crystals, and indicates the evolution of antiferromagnetic long-range order in our films at the lowest temperature.

Such a steep change in $R_H$ is not observed in Se-substituted samples. $R_H$ values at 300 K is not different much among Fe(Se$_{0.4}$Te$_{0.6}$) and Fe(Se$_{0.5}$Te$_{0.5}$) thin films, which suggests that the isovalent substitution of Se to Te does not explicitly dope carriers to FeTe. $R_H$ exhibits a slight decrease from 300K to 50 K in contrast to that of FeTe. Fe(Se$_{0.4}$Te$_{0.6}$) exhibits a sign reversal below 50K, whereas Fe(Se$_{0.5}$Te$_{0.5}$) does not. In our previous study [9], we suggested that the negative $R_H$ just above $T_c$ may be the signature of higher $T_c$ superconductivity. However, the present result means that the low-temperature downturn in $R_H$ is not a necessary condition for higher $T_c$. Fe(Se$_{0.4}$Te$_{0.6}$) shows a sign reversal from positive to negative while its $T_c$ is lower than that of Fe(Se$_{0.5}$Te$_{0.5}$). Therefore, the condition for higher-$T_c$ should be much more complicated, and we need to perform further analysis to extract an essential factor to determine $T_c$ of Fe*Ch*.

4. Discussion

One of the ways to shed light on the interplay of electrons and holes is to calculate the mobility of carriers. We thus try to extract a mobility value for each electron and hole channel using a semiclassical two-band Drude model. Fortunately, the carrier density has been already calculated for FeTe both at antiferromagnetic and nonmagnetic states [17]. Thus, we can use these values. In the semiclassical two-band model, $\rho$ and $R_H$ are described as $\rho = e^{-1}(\mu_h n_h + \mu_e n_e)^{-1}$ and $R_H = e^{-1}(\mu_h^2 n_h - \mu_e^2 n_e)/(\mu_h n_h + \mu_e n_e)^2$, where $\mu_h$, $\mu_e$, $n_h$, and $n_e$ are hole mobility, electron mobility, hole density and electron density, respectively. Once $n_h$ and $n_e$ are given, we can calculate $\mu_h$ and $\mu_e$ from the measured $\rho$ and $R_H$. Figure 6(a) shows the temperature dependence of $\mu_h$ and $\mu_e$. Below 70 K, we used the values of $n_h$ and $n_e$ for the antiferromagnetic state, while above 70 K we used the values for the nonmagnetic state corresponding to the situation of a compensated metal [17]. However, we should be careful when discussing the data above 70 K, because the high-temperature phase of FeTe is not a nonmagnetic state. The magnetic susceptibility measurements clearly indicate the paramagnetic behavior above 70 K [12]. Therefore, the values of $n_h$ and $n_e$ that we used above 70 K are probably overestimated [18], and the calculated $\mu_h$ and $\mu_e$ are underestimated. In case of Fe(Se$_{0.4}$Te$_{0.6}$) and Fe(Se$_{0.5}$Te$_{0.5}$), we can find no calculation of carrier density for these particular compositions, so that again we used the values of FeTe at the nonmagnetic state [19].

Figure 6 summarizes the temperature dependence of carrier mobility. For FeTe, $\mu_h$ is larger than $\mu_e$ at high temperature, indicating that hole channel dominates the normal-state transport. It should be noted that $\mu_e$ is close to zero, suggesting almost no contribution of the electron channel to electric transport. The hole-channel conduction is replaced by the electron-channel one at low temperatures. What is striking is that the mobility of holes is suppressed almost to zero in FeTe even though the band calculation predicts that a plenty of holes still exist in the antiferromagnetic state [17]. This can be understood as the localization of holes at low temperatures.

The results for Fe(Se$_{0.4}$Te$_{0.6}$) and Fe(Se$_{0.5}$Te$_{0.5}$) are shown in Figs. 6(b) and 6(c). We can see again that $\mu_h$ is larger than $\mu_e$ at high temperature as is observed in FeTe. However, $\mu_e$ is larger than that of FeTe. In particular for Fe(Se$_{0.4}$Te$_{0.6}$), $\mu_e$ exceeds $\mu_h$ below 50 K without antiferromagnetic transition, which means that electrons and holes coexist even at low temperature in Se-substituted samples. Thus we infer that the itinerancy of both electron and hole channels is necessary for the occurrence of superconductivity in Fe*Ch*. In case of Fe(Se$_{0.5}$Te$_{0.5}$), the evolution of $\mu_e$ seems to be slightly weaker than Fe(Se$_{0.4}$Te$_{0.6}$), but is observed again down to 40 K, and $\mu_e$ is still comparable to $\mu_h$ just above $T_c$.

The present result strongly indicates that the coexistence of itinerant electrons and holes is necessary for the occurrence of superconductivity. In other words, superconductivity does not solely occur in either an electron or a hole channel. This means the importance of inter-band scattering for the formation of Cooper pair in Fe*Ch*, which is consistent with the proposed pairing model of the sign-reversed s-wave symmetry [20, 21] as was supported by the recent STS analysis done by Hanaguri *et al.* [22]. However, the present data do not exclude the sign-preserved s-wave symmetry scenario [23], and further study is needed to clarify which is favorable pairing symmetry for the superconductivity of Fe*Ch*.

## 5. Conclusion

We have performed the Hall measurements in detail for FeTe and Fe(Se$_{1-x}$Te$_x$) thin films. The Hall coefficient ($R_H$) of FeTe exhibits a sign reversal from positive to negative across the transition from the paramagnetic to antiferromagnetic states, indicating a drastic reconstruction of the band structure. The electric transport looks to be dominated by electrons in the antiferromagnetic state. With increasing Se substitution, the mobility of electrons increases, and sometimes exceeds that of holes exhibiting a different type of sign reversal in $R_H$. This is a robust precursor to the occurrence of superconductivity, and suggests an importance of inter-band scattering for the pairing state.

We thank Zhong-Yi Lu and Xiao Tang for fruitful discussion, and Ryo Tanaka for technical assistance in the early stage of study.

Table I Sample specifications

| Composition | name | Substrate | $c$ [Å] | Thickness [nm] | $T_{c0}$ [K] |
|---|---|---|---|---|---|
| FeTe | FeTe-1 | MgO(100) | 6.285 | 165 | -- |
| FeTe | FeTe-2 | LaAlO$_3$(100) | 6.275 | 165 | -- |
| Fe(Se$_{0.4}$Te$_{0.6}$) | S40T60-1 | MgO(100) | 5.891 | 11 | 6.0 |
| Fe(Se$_{0.4}$Te$_{0.6}$) | S40T60-2 | LaAlO$_3$(100) | 5.910 | 11 | 5.6 |
| Fe(Se$_{0.5}$Te$_{0.5}$) | S50T50-1 | MgO(100) | 5.904 | 200 | 10.0 |
| Fe(Se$_{0.5}$Te$_{0.5}$) | S50T50-2 | LaAlO$_3$(100) | 5.901 | 210 | 11.4 |
| Fe(Se$_{0.5}$Te$_{0.5}$) | S50T50-3 | LaAlO$_3$(100) | --- | 90 | 3.4 |

Figure Captions

Figure 1: Photograph of six-terminal shape sample (FeTe-1). The Au-wire leads of this particular configuration are for Hall-effect measurements.

Figure 2: X-ray diffraction of (a) FeTe thin films on MgO (100) and LaAlO$_3$ (100), (b) Fe(Se$_{0.4}$Te$_{0.6}$) thin films on MgO (100) and LaAlO$_3$ (100), (c) Fe(Se$_{0.5}$Te$_{0.5}$) thin films on MgO (100) and LaAlO$_3$ (100).

Figure 3: Transmission electron microscopy (TEM) images of (a) FeTe, and (b) Fe(Se$_{0.5}$Te$_{0.5}$) thin films. In all pictures, the left side is a substrate, and the right side is the Fe*Ch* films.

Figure 4: Temperature dependence of the resistivity of FeTe, Fe(Se$_{0.4}$Te$_{0.6}$), and Fe(Se$_{0.5}$Te$_{0.5}$) thin films.

Figure 5: Field dependence of Hall resistivity of (a) FeTe and (b) Fe(Se$_{0.5}$Te$_{0.5}$) thin films. (c) Temperature dependence of Hall coefficients of FeTe, Fe(Se$_{0.4}$Te$_{0.6}$), and Fe(Se$_{0.5}$Te$_{0.5}$) thin films.

Figure 6: Temperature dependence of the mobilities of electrons and holes in (a) FeTe, (b) Fe(Se$_{0.4}$Te$_{0.6}$), and (c) Fe(Se$_{0.5}$Te$_{0.5}$).

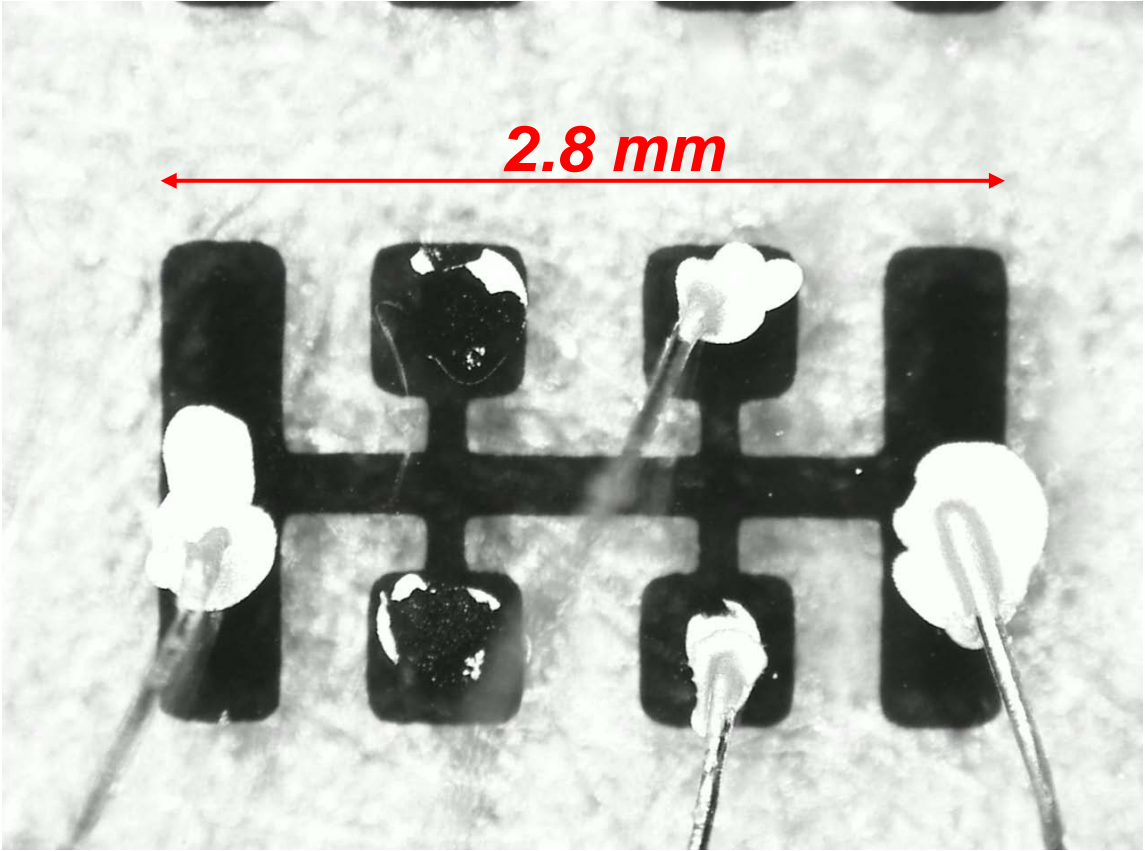

Figure 1

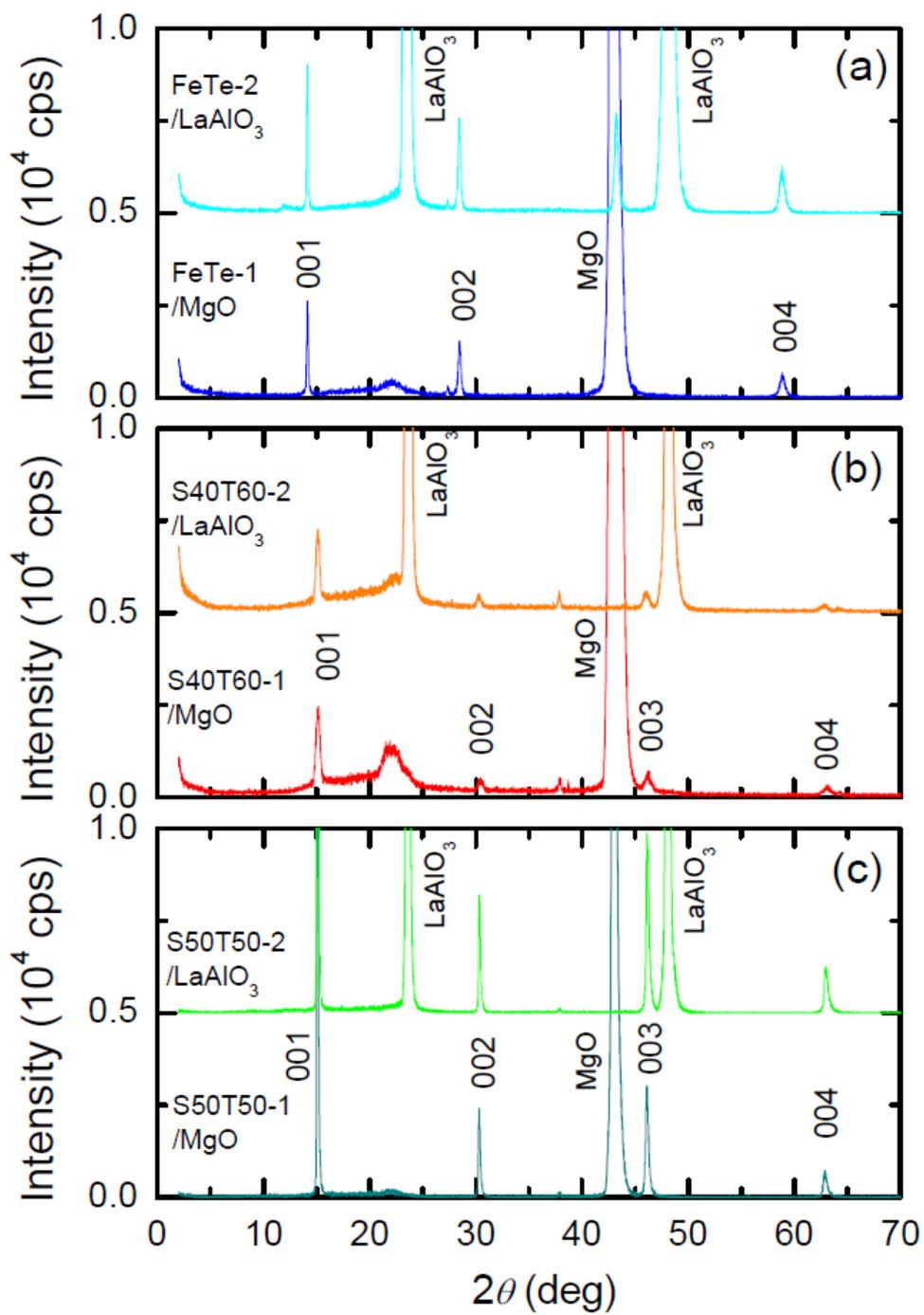

Figure 2

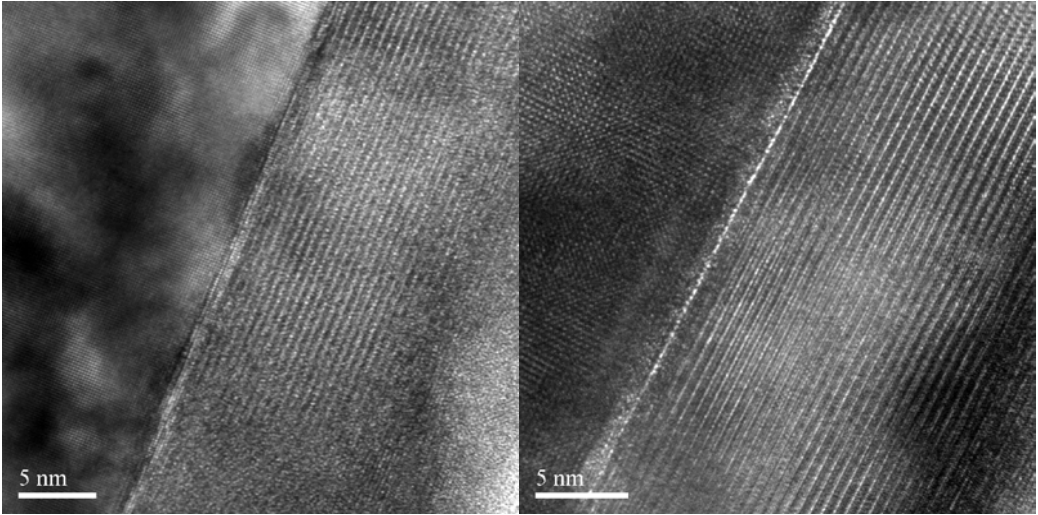

(a)   FeTe-1 (on MgO)         FeTe-2 (on LaAlO$_3$)

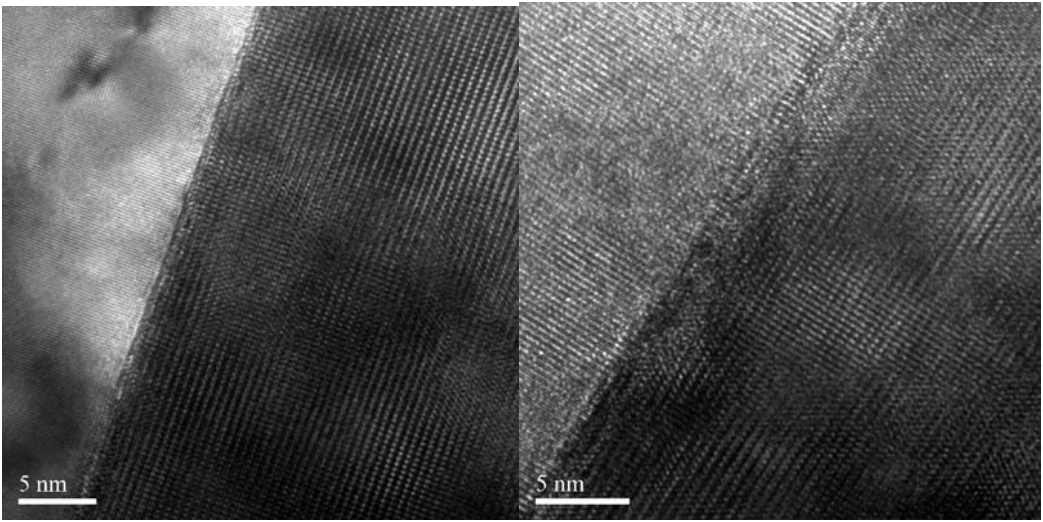

(b)   S50T50-1 (on MgO)       S50T50-2 (on LaAlO$_3$)

Figure 3

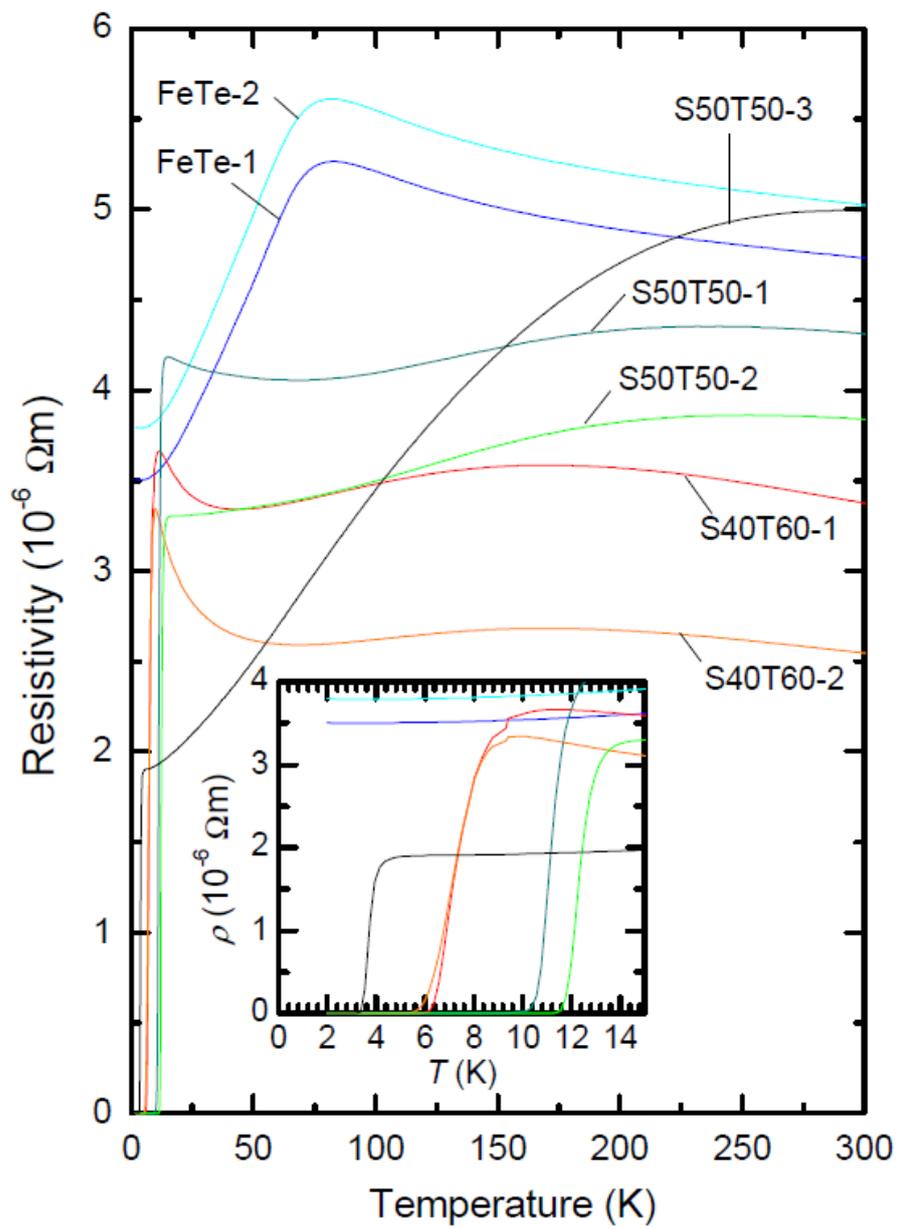

Figure 4

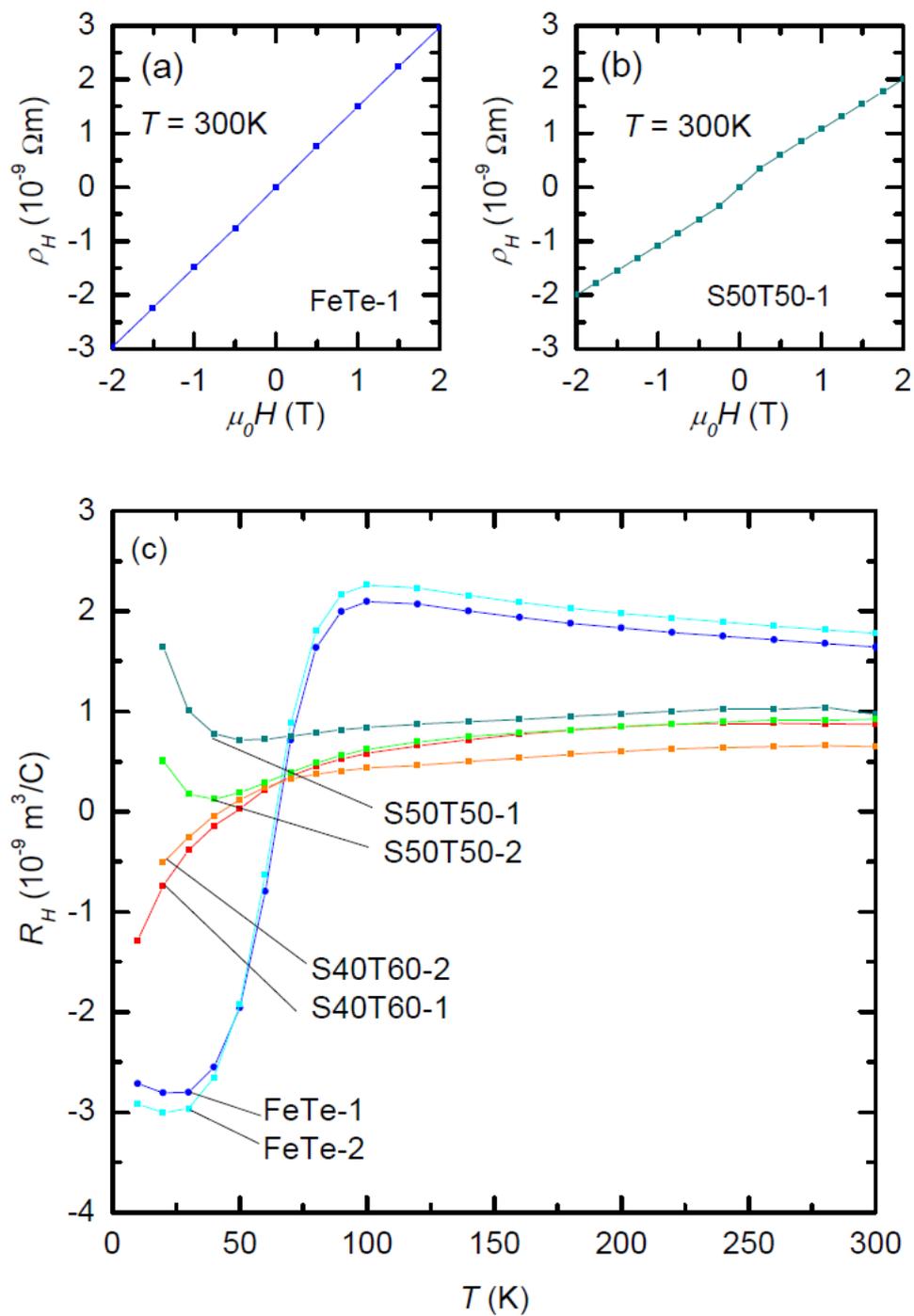

Figure 5

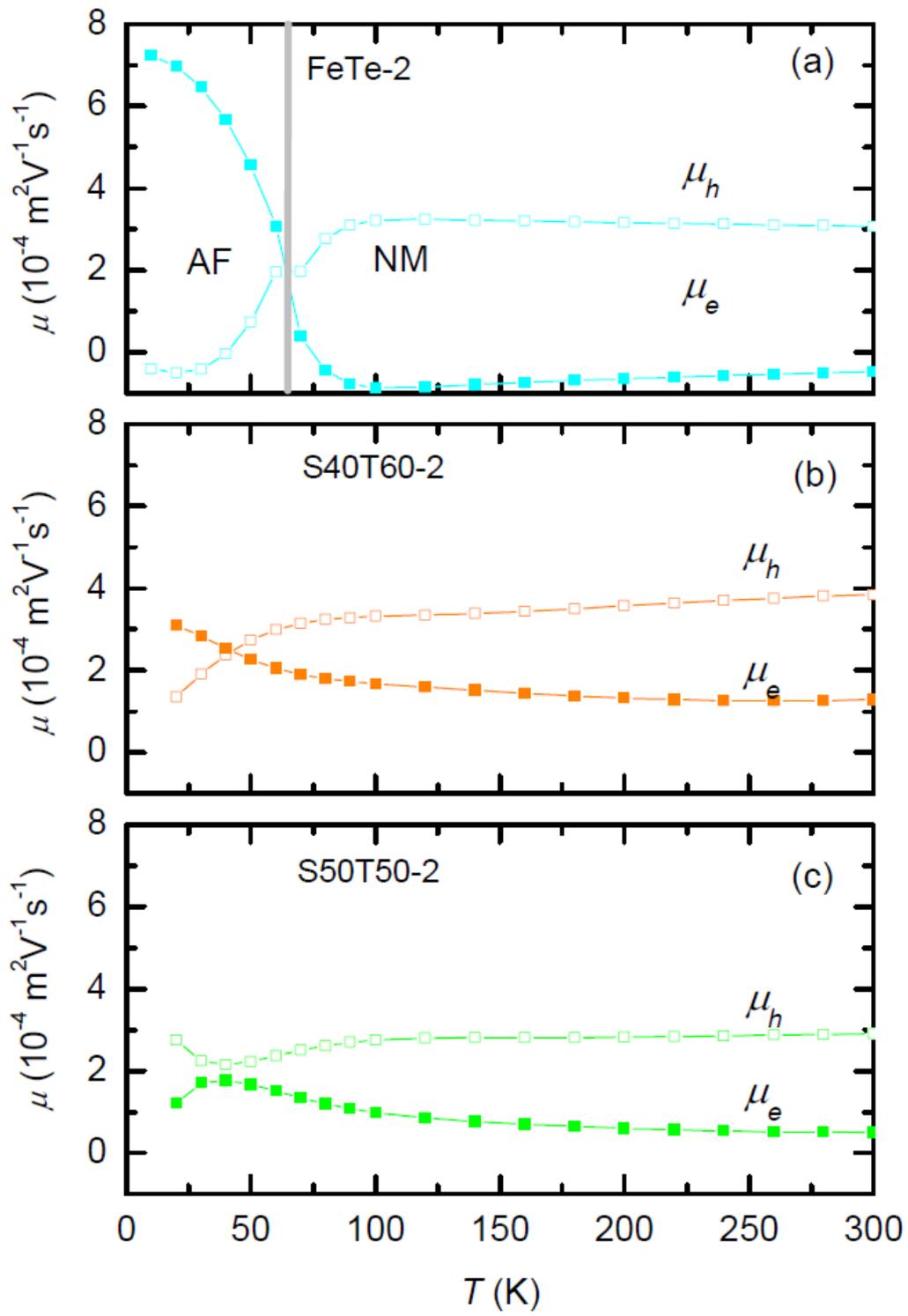

Figure 6